\documentclass[aip,graphicx]{revtex4-1}
\usepackage[pdftex]{graphicx}
\usepackage{bm}
\usepackage{dcolumn}

\draft 

\begin{document}

\title{A complex network approach to robustness and vulnerability of spatially organized water distribution networks}

\author{Alireza Yazdani}
\email[Corresponding author:]{a.yazdani@cranfield.ac.uk}
\author{Paul Jeffrey}
\affiliation{School of Applied Sciences, Cranfield University, MK43 0AL, UK}


\date{\today}

\begin{abstract}

In this work, water distribution systems are regarded as large sparse planar graphs with complex network characteristics and the relationship between important topological features of the network (i.e. structural robustness and loop redundancy) and system resilience, viewed as the antonym to structural vulnerability, are assessed. Deterministic techniques from complex networks and spectral graph theory are utilized to quantify well-connectedness and estimate loop redundancy in the studied benchmark networks. By using graph connectivity and expansion properties, system robustness against node/link failures and isolation of the demand nodes from the source(s) are assessed and network tolerance against random failures and targeted attacks on their bridges and cut sets are analyzed. Among other measurements, two metrics of meshed-ness and algebraic connectivity are proposed as candidates for quantification of redundancy and robustness, respectively, in optimization design models. A brief discussion on the scope and limitations of the provided measurements in the analysis of operational reliability of water distribution systems is presented.

\end{abstract}


\maketitle 


\section{Introduction}
The resistance and vulnerability of infrastructure systems exposed to perturbations and hazards is becoming a more important consideration during the design and operation of such systems. Water distribution systems are among the most critical infrastructures which are subject to numerous perturbations ranging from the common failures (e.g. breakdown of the ageing pipes and pumping units causing low pressure and disruptions to supply) to the security risks and exposure of the networks to natural or man-made disasters (e.g. earthquakes, flooding, vandalism, etc). The ability of the system to resist, mitigate and overcome stresses and failures and their consequences is consequently receiving increased attention from the managers and system engineers with robustness and resilience gradually becoming very important considerations along with the more typical considerations of low cost and reliable operation. Moreover, proper understanding of the vulnerable and critical locations in distribution systems provides invaluable information that may be used to inform asset management practices and rehabilitation programs leading to more realistic risk assessments and the development of defensive strategies to ensure network survival in the case of extreme events and natural or man-made disasters. Reliable assessment of water distribution network (WDN) performance in the face of various stresses is dependent on close and clear definition of network characteristics such as resilience and robustness. A resilient system shows (i) Reduced failure probabilities (ii) Reduced failure consequences and (iii) Reduced time to recovery, characterized by four infrastructural qualities of robustness, redundancy, resourcefulness and rapidity (Bruneau et al., 2003) which incorporates the notions of risk (probability of failure and its consequences), reliability, recovery and system tolerance pre- and post-failure. While a comprehensive assessment of resilience to include resourcefulness and rapidity requires operational considerations and access to field data, this work is confined to quantifying structural robustness and redundancy as topological aspects of network resilience. Structural robustness is the optimal connectivity of network components which gives rise to higher tolerance against disruption of operation as a result of component failures. Redundancy, a similar indicator of network connectivity, is the presence of independent alternative paths (usually looped structures) between the source and demand nodes which can be used to satisfy supply requirements during disruption or failure of the main paths.\\

Complex networks are governed by communication and distribution laws in such a way that the behavior and interaction of the individual elements, taken together with the non-trivial network configurations, may cause the overall system performance to be much different from the sum of the performances of the individual parts. Examples of such networks range from the internet and communication networks to urban roads and power grids with typical complexities and uncertainties observed in network congestions, traffic jams or power cuts. Analysis of the vulnerability of complex networks using graph theory techniques is concerned with the study of network building blocks and motifs and identification of structural weaknesses, critical locations and impacts of failures of core components and highly connected nodes known as the hubs (Albert et al. 2000). This is usually carried out using simulation techniques which assess the impacts on efficiency and performance of the network as a result of random failures or successive targeted attacks (Crucitti et al. 2005). Notable studies concerned with the structural analysis and assessment of the vulnerability in complex networks among other sectors include resilience of urban transport networks (Zio and Sansavini 2007, Masucci et al. 2009), vulnerability of power grids (Bompard et. al 2009, Crucitti et al. 2005 and Holmgren 2006) and the World Wide Web (Albert et al. 1999). Furthermore, other methodologies are proposed that analyze the scenarios of cascading failures in the networks, along with the traditional path-based methods in the analysis of infrastructure network reliability (Dueñas-Osorio and Vemuru, 2009). \\

Despite the extensive use of graph models in the analysis of pipe networks (Kesavan and Chandrashekar 1972, Gupta and Prasad 2000 and Deuerlein 2008) vulnerability and robustness of WDNs has not been systematically exposed to the analyses by graph theory and complex network techniques. Among the few works in this area, are the work by Jacobs and Goulter (1988, 1989) who showed that networks that are most invulnerable to failures are regular graphs with equal number of links incident to each node, while the inverse relationship is not necessarily true due to the existence of bridges (links whose removal disconnects the network) and articulation points (nodes whose removal along with the removal of their incident links disconnects the network). Kessler et al. (1990) used graph theory to develop a methodology for least-cost design of invulnerable WDNs by incorporating reliability in the design of the network. Ostfeld and Shamir (1996) and Ostfeld (2005) utilized graphs to study the selection of one-level system redundancy "backups" in a WDN undergoing failure. However, the use of purely topological graph theory in reliability analysis of WDNs was shown to have limited scope. Walski (1993) showed that link-node representation does not account for the importance of the valves in a WDN reliability analysis. Ostfeld and Shamir (1993) classified failures into two types of: (1) failures of system components (2) failures in meeting consumers’ demands, and emphasized that "these two types of failures should not be considered separately" as they are strongly connected. The analysis of the connectivity, reliability and risk-based design of WDNs, on the other hand, have been extensively dealt with using simulations, optimization algorithms and non-deterministic techniques (Ostfeld 2004, Babayan et al. 2005 and Kapelan et al. 2006). In general, network reliability deals with the assessment of probability of link or node failure while vulnerability, viewed here as the antonym to robustness, is more concerned with the impacts of failed components or subsystems. While a thorough reliability analysis may not be viable by using purely topological metrics only, structural vulnerability and robustness of WDNs may still be studied within such a framework. Moreover, as demonstrated later, recent advancements in the field of complex networks in terms of the available network measurements, prove invaluable towards quantification of structural properties of distributed systems and understanding robustness and redundancy, as two topological components of reliability and the level of available service to the consumers, that in turn may be used as parameters within the framework of design models to optimize network reliability against size and cost. However, given that such topological analysis of water distribution networks does not take into account the operational specifications of water systems and the level of service provided to consumers, it may only be regarded as an assessment of the "necessary conditions" (Ostfeld, 2001) for supplying required demand for water.\\

The aim of this work is to provide more insight towards resilience or vulnerability of water distribution systems by looking at the connectivity patterns and network configurations. In particular, techniques from "Complex Networks" and "Graph Theory" are employed to estimate redundancy and quantify robustness against node/link failures in WDNs. A WDN with multiple interconnected elements is represented as a link-node planar graph which comprises of nodes for features at specific locations and links which define the relationship between such nodes. It is common to represent pipes by links and junctions such as pipe intersections, reservoirs and consumers by graph nodes. With such representations, WDNs resemble complex networks where network reliability and the severity of disruption as a result of failures depend on network layout and structure of the cycles and loops as alternative supply paths. The key observation here is that "structure affects function" (Strogatz 2001) as network architecture reveals important information on robustness, vulnerability and possible operational consequences as a result of component failures. The present analysis utilizes spectral graph theory and statistical measurement of complex networks to study the structure of some benchmark WDNs in relation to robustness and vulnerability and quantify robustness and redundancy as the two topological aspects of network resilience. Basic connectivity metrics, spectral gap and algebraic connectivity along with the statistical measurements such as  clustering coefficient, meshed-ness, central-point dominance and degree distribution are deployed to signify features related to graph robustness and strength such as bottlenecks, structural holes and cut vertices. Such intuitive and relatively easy to implement methodology reveals important information on structural vulnerability and resilience to perturbations in water networks. Consequently, topological assessment, ideally, may be used as a decision-making support methodology in relation to the design and construction of robust and optimally-connected next-generation of WDNs.\\

\section{NETWORK MODELS AND STRUCTURAL METRICS}
The theory of complex networks employs techniques from graph theory and statistical physics to classify different network models, to analyze their structural complexities and quantify the vulnerability, robustness and error and attack tolerance of the networks (Albert et al. 2000). Complex networks typically belong to one of the groups of: technological, biological, social or information networks (Newman, 2003) with their topological structures categorized as: centralized, decentralized and distributed depending on the underlying hierarchical or redundant configurations. Selection of a relevant set of statistical measurements to analyze network robustness is, however, largely determined by the fact that WDNs are spatial technological networks and need to be treated as planar graphs. Throughout, a network is represented as a mathematical graph $G=G(V,E)$ in which $V$ is the set of all graph nodes with $n$ elements and $E$ is the set of graph edges with $m$ elements. Table (I) provides a brief definition of some of the measurements related to network robustness and redundancy together with an indication of which water distribution network characteristic is being assessed through each metric. Water distribution systems could be potentially weighted bi-directional networks, given the pipes and nodes physical attributes and the possibility of flow redirection which may take place as a result of the failures and isolation of network parts. In general, the direction of network flow and weighting allocated to graph links or nodes is determined by physical and operational parameters and the costs associated with water supply in networks under study for which no data was available and hence the studied networks are treated as undirected and unweighted graphs. Moreover, WDNs in this analysis are assumed to be planar graphs so that graph edges intersect only at a node mutually incident with them (Chartrand and Lesniak, 1996). \\
\begin{table}[]\tiny
\caption{Summary of structural metrics used to assess redundancy and robustness of networks}
\begin{ruledtabular}
\begin{tabular}{|p{2.8cm}|p{6cm}|p{3.4cm}|p{2cm}|p{1.5cm}|}

\bf{Metric}                        & \bf{ Definition}                           &\bf{Equation}                     & \bf{Refernce}                         & \bf{Quantifying }  \\ \hline 

Link density&The fraction between the total and the maximum number of links& $q=\frac{2m}{n(n-1)}$&	Jamakovic and Uhlig (2007)&Redundancy\\\hline
Average node-degree&	Average value of the node-degree distribution&	$<k>=\frac{2m}{n}$&	Costa et al. (2007)	& Robustness\\\hline
Meshed-ness	& The fraction between the total and the maximum number of independent loops in planar graphs & $R_m=\frac{m-n-1}{2n-5}$ &Buhl et al.(2006)&	Redundancy\\\hline
Diameter&The maximum geodesic length of the shortest path between all pairs of nodes&$d=max(d_{ij})$&Costa et al. (2007)&Robustness\\\hline
Average path length&Average value of the geodesic distances between all pairs of nodes&	$l=\frac{1}{n(n-1)}\Sigma_{(i≠j)} d_{ij}$ &Costa et al. (2007)&Robustness\\\hline
Clustering coefficient&The fraction between the total triangles and the total connected triples&$C=\frac{3N_\Delta}{N_3}$ &Wasserman and Faust (1994)&Redundancy\\\hline
Central-point dominance& Average difference in betweenness of the most central point and all others&$C_B=\frac{1}{(n-1)}\Sigma_i(B_{max}-B_i)$ &Freeman(1997)&Robustness\\\hline
Spectral gap&	The difference between first and second eigenvalues of graph's adjacency matrix& $\Delta \lambda$ &Estrada (2006)& Robustness\\\hline
Algebraic connectivity&	The second smallest eigenvalue of Laplacian matrix of the network& $\lambda_2$ &Fiedler (1973)&Robustness\\
\end{tabular}
\end{ruledtabular}
\end{table}

The metrics introduced are mainly divided into two groups of statistical and spectral measurements. Statistical measurements are those which quantify organizational properties of the network based on the most frequent motifs and structural patterns and relate them to network robustness and the dynamics on network. Such measurements range from more basic metrics such as network size n, network order m, link density, number of independent loops given as $N_l=m-n+1$ and graph diameter $d$, to other statistical metrics including degree distribution and average node-degree $<k>$, meshed-ness for planar graphs defined as $R_m$ , clustering coefficient $C$ and central-point dominance $C_B$ as a measure of relative betweenness centrality or network centralization. Spectral metrics derived from the spectrum of network adjacency matrix, quantify network invariants that, taken along with the described statistical measurements, reveal useful information on well-connectedness of the network, connectivity strength and failure tolerance. Two basic yet important such connectivity metrics are link-connectivity and node-connectivity (Kessler et al. 1990) which quantify the minimum number of attacks or failures required to render a group of the nodes disconnected. In general, the value of node-connectivity is smaller than link-connectivity which could be interpreted that the existence of unreliable network nodes may have more severe consequences in terms of the failures and recovery than the existence of unreliable network links. However, these measurements may become trivial in WDNs as a result of low redundancy and sparseness at transmission or sub-urban levels and due to single-connections to which supply water to the end-users. Therefore, alternative connectivity measures such as "spectral gap" and "algebraic connectivity" (Table I) are employed in this analysis in order to quantify well-connectedness and strength of the network connections. Using above graph representations, the Laplacian (or Kirchhoff) matrix of graph $G$ with $n$ nodes is a $n×n$ matrix $L=D-A$ where $D=diag(d_i)$ and $d_i$ is the degree of node $i$ and $A=(a_{ij})$ is the adjacency matrix of $G$ where $a_{ij}=1$ if there is a link between nodes $i$ and $j$ and $a_{ij}=0$ otherwis. Spectral gap $\Delta$ is the difference between first and second eigenvalues of adjacency matrix $A$ and is used to identify Good Expansion (GE) properties in graphs. GE networks are those with their topology of network-connections structured such that "any set of vertices connects in a robust way to other nodes, even if the graph is sparse" (Donneti et. al, 2006). Non-GE networks are those with the presence of bottlenecks, articulation points or bridges and are easily split and isolated into two or more parts by removing those nodes or links. A necessary condition for GE property is that the spectral gap is sufficiently large (Estrada, 2006) and hence a small spectral gap may indicate the lack of GE properties. Another spectral measurement quantifying the strength of network connections is the algebraic connectivity, defined as the second smallest eigen-value of graph Laplacian matrix $L$ Fiedler (1973) and extensively discussed in Mohar (1991), Ghosh and Boyd (2006) and Jamakovic and Uhlig (2007) for its properties and applications towards analysis of graph robustness to node and link failures. While the smallest eigenvalue of graph Laplacian is zero and its multiplicity equals the number of connected components in a network, larger value of algebraic connectivity indicates the fault tolerance of the network and its robustness against efforts to cut it into parts. In the following section, statistical and spectral measurements for the studied benchmark water distribution models are derived and their relationship to network robustness and vulnerability is interpreted.\\

\section{CASE STUDIES AND NUMERICAL ANALYSIS}
Four benchmark water distribution networks are studied and their structural properties explored. These networks, listed below, are among the models reported in the literature with their data-files accessible online at the University of Exeter Centre for Water Systems (CWS Benchmarks, 2007) and depicted in the Figure (1). The selected networks include two real-world and two hypothetical models chosen from a set of networks with different sizes and structures proving some diversity in terms of the network structures and data analysis. 


\begin{figure}[]
\centering
\includegraphics[width=0.7\textwidth, height=12cm]{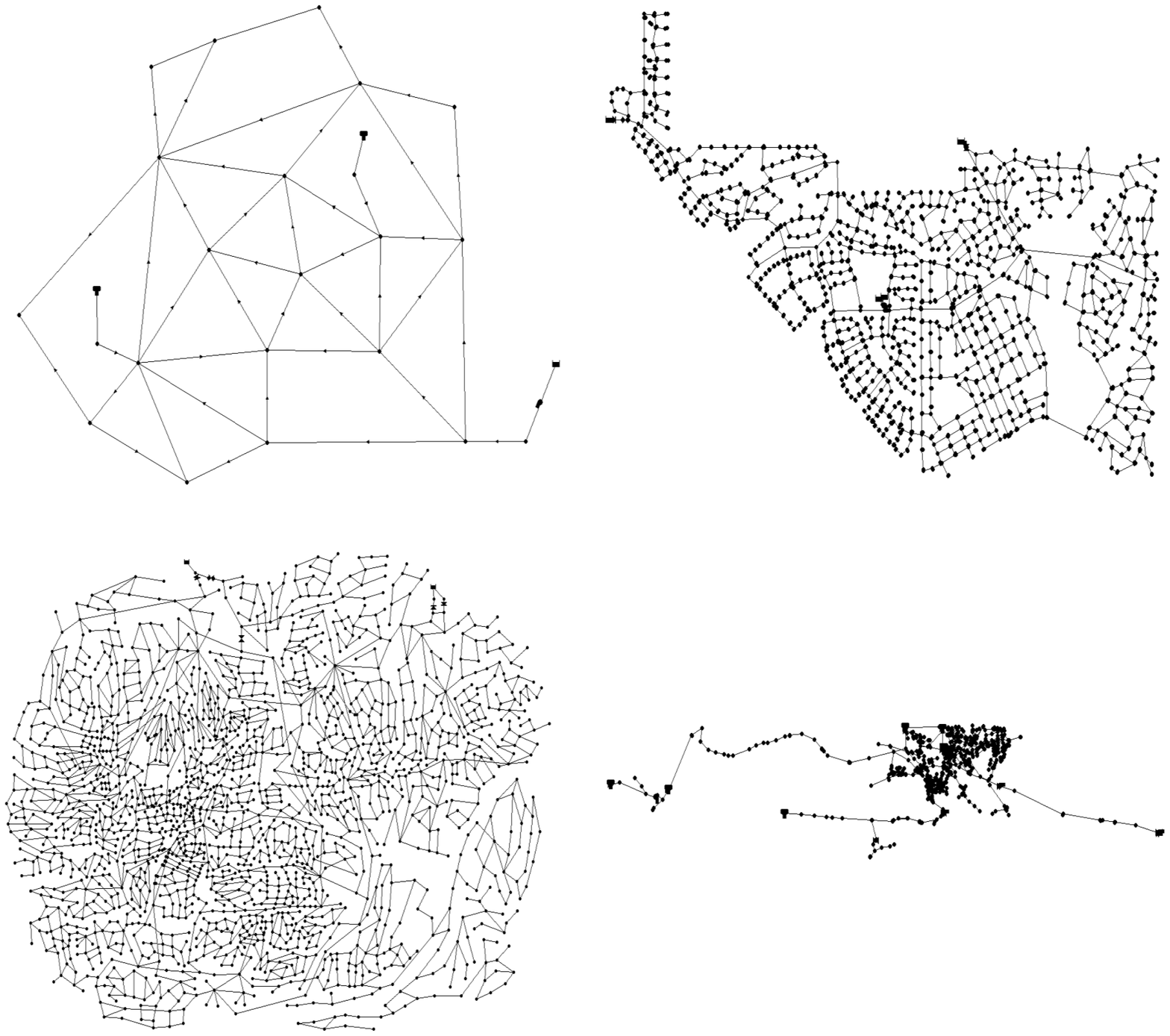}
\caption{Total network graph view of studied water distribution networks; Anytown (top-left), Colorado Springs Utilities (top-right), EXENET (bottom-left), Richmond (bottom-right).}
\end{figure}

1.	The "Anytown" design example proposed by Walski et al. (1987); a small hypothetical network with nearly regular lattice-like structure. The particular all-channel structure entails high redundancy which in turn improves network connectivity and makes the network failure-tolerant. Nodes with degrees five and two are the most frequent types of connection and low graph diameter and the dominance of triangular paths are the main characteristics of the network.

2.	The "Colorado Springs Utilities" reported in Lippai (2005); a large real sparse network with mesh-like structures and redundant rectangular loops more frequent than triangular configurations at local distribution levels. However, structural holes are viewed in the network profile at a more global view which suggests that large segments of the network can be disconnected by failure of a relatively small number of nodes or links.

3.	The "EXNET" proposed by Farmani et al. (2004); a very large hypothetical sparse network that enjoys redundant triangular and trapezoidal configurations. Despite the existence of a few highly connected nodes (hubs) which leave the network vulnerable to attacks and random failures of these hubs, decentralized network structure and high redundancy may help to improve robustness and network efficiency under normal circumstances or in the case of failures. 

4.	The "Richmond" example from UK Yorkshire Water reported in Van Zyl et al. (2004); a medium size real sparse network with high link density in the urban centre and a very sparse configuration elsewhere. This network has an unusual geographic spread where given the low loop redundancy in the suburban areas (where the reservoirs and water supply sources are located) may result in large-scale disruptions as a result of failure of pipes or the outage of system component at network upstream. This specific network layout indicates the existence of geographical and urban constraints and that this water distribution network has evolved and expanded over a longer period of time than as an outcome of a single optimized construction.
Network analysis is undertaken using the open-source graph manipulation software "igraph" (Csardi and Nepusz, 2006) which can be used as an extension package on statistical computing software "R" (R Development Core Team, 2009) available free online. Calculations of the network measurements take place in few simple steps listed below: 

1.	Install and call igraph library package in “R” environment (other library packages such as "matrix", "stats" or "graph" might be required depending on the undertaken analysis).

2.	Read the network file, structured in a format supported by igraph functions for reading external files.  

3.	For each measurement use the associated functions, directly or in combination with others, by using "R" commands.

4.	Most of the calculations are performed in linear or polynomial computation time depending on the used function and network size (Csardi and Nepusz, igraph Reference Manual). 

Some of the basic topological and statistical measurements of these benchmark models are summarized in Table (II). \\
\begin{table*}[b]
 \caption{Basic network measurements for the benchmark water networks ($n$ = nodes, $m$ = links, $q$ = link density, $M$ = maximum node-degree, $N_l$ = number of independent loops, $<k>$ =average node-degree, $d$ = diameter, $l$ = average path-length, $C_B$ = central-Point dominance)}
\begin{ruledtabular}
\begin{tabular}{ccccccccccc}

Network                             & Network Type         &  $n$         &    $m$          &  $q$                 &  $M$       &$N_l$     & $<k>$          & $d$         &$l$         & $C_B$     \\ \hline 
Anytown                             & Hypothetical             &   25           & 44             &   14.67 \%       & 7              &20              &  3.52              &7           & 2.89     &   0.23       \\ 
Colorado Springs             &  Real                           &   1786     & 1994           & 0.13 \%          & 4              &209              &  2.23            &69         & 25.94   &   0.42       \\ 
EXNET                                &  Hypothetical            & 1893       & 2467	      &  0.14 \%         &11            & 575               & 2.60	        &54      &20.60    &  0.28       \\
Richmond                          &  Hypothetical            &  872	     &  957	              &   0.25 \%         & 4            & 86               & 2.19	        &135      &51.44    &  0.56       \\

\end{tabular}
\end{ruledtabular}
\end{table*}

Numerical calculations presented in Table (II) relate to the following observations: Firstly, the studied WDNs are sparse networks with very low values of link density (Table II). The "Anytown" network has a relatively higher link density while still regarded a sparse network. It is observed that high link-density does not necessarily imply high level of loop redundancy, particularly in tree-like networks spread linearly against the plane (e.g. "Richmond"). Secondly, the average node degree in the studied WDNs is particularly low and cumulative degree distribution of these networks follow exponential or uniform trends (Figure 2), which may have implications on robustness or well-conceived the network layouts are, viewed along with the expansion properties discussed in the next section. The studied networks are to a large extent (though not strictly) planar graphs where most of the graph links intersect at their endpoints. In such networks the graph structures usually range between simple tree-like networks (corresponding to $<k> =2$) and two dimensional regular lattices (corresponding to $<k> =4$). Therefore, the number of independent loops $N_l$ (Table 2) and estimated network meshed-ness (Table III) as defined by Buhl et al. (2006) can be used to evaluate the local structural properties and redundancy in the network. Finally, clustering coefficient values (Table III) show some degree of correlation with average meshed-ness. This is because the clustering coefficient provides a measurement of the density of triangles among all other types of building blocks. However, due to the frequency of non-triangular loops in the studied networks, extra care should be taken to not to use clustering coefficient as an ultimate measure of redundancy since it only provides estimation for the density of the transitive triangles rather than other looped motifs. Numerical measurements based on graph adjacency matrix spectrum to quantify network robustness are given in Table (III).\\


\begin{figure}[t]
\centering
\includegraphics[width=0.8\textwidth, height=10cm]{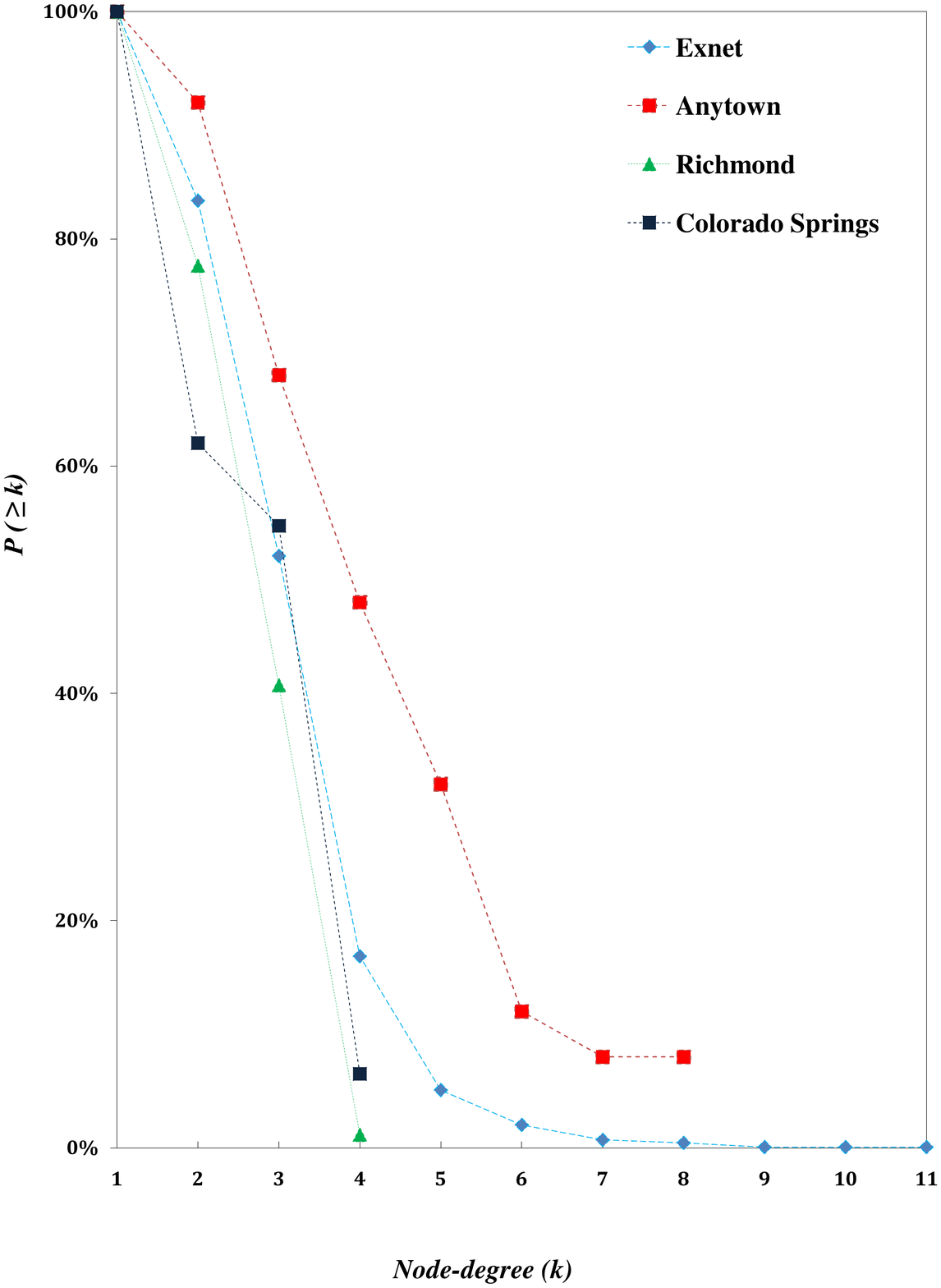}
\caption{The GIS view (top) and the graph representation (bottom) of the two of the four water distribution networks; "Colorado Springs" (left) and "Richmond" (right).}
\end{figure}

    
\begin{table}[t]
 \caption{ Redundancy and robustness quantifications for the benchmark water networks ($R_m$ = meshed-ness, $C$ = clustering coefficient,  $\Delta \lambda$ = spectral gap, $\lambda_2 $= algebraic Connectivity)}
\begin{ruledtabular}
\begin{tabular}{ccccc}
Network                        &       $R_m$                          & $C$                      & $\Delta \lambda$                  & $\lambda_2$      \\  \hline
Anytown                       &         $44.44 \%$               &    0.32                 & $1.51$                                       &  0.2877            \\
Colorado Springs       &         $5.85\%$                  &      0.0008           & $0.03$                                       &  0.0005            \\
Kumasi                          &         $15.21\%$               &    0.0517              &  $0.12$                                     &  0.0011            \\
Richmond                     &         $4.95\%$                  &       0.0402          &   $0.07$                                     &  0.0001               \\ 
\end{tabular}
\end{ruledtabular}
\end{table}
\section{Discussion}
Findings on structural and statistical properties of studied WDNs are comparable to the reported measurements for other technological networks (see Newman, 2003, p11). Technological networks are usually designed and expanded to provide the most efficient distribution of flow and to minimize the associated costs. In the case of civil infrastructure assets identified by geographical organizations and spatial locations of the nodes and links, a typical network hierarchy has a sparse topology at transmission levels with less sparse mesh-like structure at the distribution levels in urban centers. In WDNs, this is observed as few long-distance high-capacity trunk mains at transmission level which carry water from the source(s) to the urban sub-networks that are mostly made of the grid-like structures of small distribution pipes with certain loop redundancy provisioned through active or standby alternative supply routes. \\

The average-node degree in a WDN is generally low, mostly due to the physical constraints (represented through network graphs that are planar and homogeneous), which in turn implies that a majority of the nodes or links have more or less equal importance and failure consequences only as far as purely topological degree-related metrics are concerned. The exception would however be the cut-vertices and cut-edges whose failures may decompose the network and result in isolation of the parts from the water supply sources. This approach may (and will) totally change, once considering characteristics such as the pipe size and nodal demand instead of nodal degree or betweenness. The geographical spread of WDNs is determined by the Euclidean distances between the nodes and the associated costs, in addition to the constraints of urban roads and other constructions, which in turn prevents the formation of highly connected hubs and centralized network structures. Vulnerability analysis of complex networks is largely concerned with the operational consequences of single or multiple failures as a result of incidents and attacks on the hubs or the most central elements exercised through scenarios of the removal of nodes (links) randomly or selectively in decreasing order of node degrees. In the absence of the hubs in most of real WDNs, a more realistic analysis should be focused on the other important structural features such as bridges and bottlenecks and their relationship with GE properties in the network. It has been found (Estrada, 2006) that complex networks with "simultaneous existence of GE and uniform degree distribution" are more robust against the failure of their nodes and links. The studied WDNs are found (Figure 2) to be single-scale networks, as opposed to several scale-free models reported for non-technological networks (Albert et al. 2000). This is to say that the graph degree-sequence for such networks is either uniformly distributed (such as in regular graphs or perfect grids) or it belongs to the exponential family with a pronounced peak with typically up to a maximum of four connections per junction in real water distribution networks. \\

It is impossible or very unrealistic to use a single metric to characterize network structures or capture a lot of information on different aspects of graph robustness and vulnerability. Therefore being able to directly compare and rank vulnerability of networks is another function-specified problem which depends on the existence of a conclusive set of measurements followed by heuristic interpretations. However, based on the above observations, it is possible to simplify the task of the assessment of structural vulnerability or resilience to the quantitative measurements of algebraic connectivity, spectral gap and meshed-ness coefficient supported by other measurements to quantify network robustness and redundancy, respectively. Considering the algebraic connectivity, a comparison of the listed measurements shows that the hypothetical network "Anytown" is the most invulnerable against node and link failures, with "EXNET" being the second most robust network in the list. Larger value of algebraic connectivity that indicate graph tolerance to the efforts to decompose it into isolated parts, is accompanied by larger values of spectral gap representing GE properties and non-existence (or insignificance) of the network cut-sets and articulation points. Small values of graph diameter and central-point dominance quantify this property that the network nodes are mutually reachable and the network are ordered in a decentralized fashion which helps with more a more efficient and balanced distribution of flow and smaller consequences as a result of failure of the most central components. It is worth noting that both these networks are hypothetical models developed as a result of a more planned optimization process with less structural randomness than is usually found in real-world networks gradually evolved in time and understood as more static objects (Buhl et al. 2006). Understanding the vulnerability aspects of the other two (real) networks "Colorado Springs" and "Richmond" is however, not as obvious. Larger spectral gap suggests GE properties for "Richmond" along with the significance of triangular loops in the networks characterized by clustering coefficient. However, greater algebraic connectivity and meshed-ness coefficient quantify "Colorado Springs" as a more robust network with higher loop redundancy. This observation is backed up by other measurements such as slightly larger average node-degree, smaller diameter and smaller value of central-point dominance indicating that "Colorado Springs" has been ordered in a more lattice-like structure and is a less centralized better-connected network.

\section{Conclusion}

In this paper, applications of graph theory and complex network principles in the analysis of vulnerability and robustness of water distribution networks are investigated. Some benchmark water networks with different sizes and structures were studied and their vulnerability-related structural properties are quantified. These metrics, grouped as basic connectivity, spectral metrics and statistical measurements, are used to relate the network structure to resilience against failures or targeted removal of the nodes and links. This is done by taking vulnerability as the antonym to resilience and studying the two main topological aspects of resilience: robustness and redundancy. It is observed that water networks are organized as homogeneous planar graphs against two dimensional Cartesian plane and hence network vulnerability and robustness is to a great extent determined by the magnitude of algebraic connectivity and spectral gap. Other measurements are utilized to demonstrate that decentralized networks with smaller graph diameter, higher level of redundancy and more lattice-like structures are connected in a more optimal fashion, with perfect grids and regular graphs being the most invulnerable networks in this respect. The extent to which the investigated metrics can be used is elaborated and it was discussed that while each graph metric captures some information on particular aspects of network robustness and vulnerability, no unique metric can be used as the only descriptive measurement to characterize network robustness in relation to design. A thorough assessment of network vulnerability with the objective of improvement in network design will therefore require further specifications related to the physical or hydrological attributes of water distribution systems, drop in the operational concepts such as network efficiency in delivery of the utility as measured against the costs associated with the improvement in design. Such an assessment may ideally take into account the real world system specifications such as the potential differences between the network nodes to distinguish between the sinks, sources and other junctions in order to emphasize the demand for water. In addition, graph links may be allocated with weighting factors to specify physical attributes such as pipe length, capacity or the cost. Other function specific information such as geographical characteristics and flow direction to name but a few may also prove useful in a more comprehensive analysis of resilience in water distribution networks. This analysis however, as the first steps in this direction, demonstrates that techniques from graph theory and network physics are invaluable tools towards the assessment of infrastructure network robustness and better understanding their structural vulnerabilities for the practice of alternative designs. Findings in this area may be used as the guidelines for the implementation of protective measures by risk managers and will help engineers who are concerned about the design and construction of optimally-connected next-generation water distribution networks. \\



%

%


\begin{acknowledgments}
We would like to thank Leverhulme Trust for providing the financial support for this research. 
\end{acknowledgments}




\end{document}